# Strategies to Reach Ultra-High Capacitance Values for Supercapacitors: Materials Design

Ipek Deniz Yildirim[1], Ameen Uddin Ammar[1], Merve Buldu Aktürk[1], Feray Bakan[2], Emre Erdem[1,2,3,4,*]

[1]Faculty of Engineering and Natural Sciences, Sabanci University, Tuzla, 34956 Istanbul, Turkey

[2]Sabanci University Nanotechnology Research Centre (SUNUM), Tuzla, 34956 Istanbul, Turkey

[3]Center of Excellence for Functional Surfaces and Interfaces for Nano-Diagnostics (EFSUN), Sabanci University, Tuzla, 34956 Istanbul, Turkey

[4]Integrated Manufacturing Technologies Research and Application Center & Composite Technologies Center of Excellence, Sabanci University, Teknopark Istanbul, 34906 Pendik, İstanbul, Turkey

*Corresponding to* E. Erdem (E-mail: emre.erdem@sabanciuniv.edu)

**Abstract:**

This review paper highlights the recent developments in supercapacitors by pointing out the significance of appropriate electrode and device designs. We reported ten extremely high-performance supercapacitors with specific capacitance values among the highest available until now to the best of our knowledge. These state-of-the-art designs employing innovative electrode materials have been discussed along with their short descriptions. The supercapacitors collected here possess the most promising potential for facilitating next-generation systems in energy harvesting and storage. This review is just the surface that can help provide a pathway for supercapacitor research, which is still wide open for exploring and developing new advanced materials for energy applications of the future.

## 1. Introduction

Due to the increasingly diversified demands in energy storage devices, many laboratories have concentrated on energy storage research, including supercapacitors. Capacitors, which are the principal electrical circuit element, can store energy based on charge and discharge of electricity. Since the discovery of capacitors major evolutions are generated and finally supercapacitors are found. Until then supercapacitors are very popular for energy storage applications and due to its ability combine faradaic and non-faradaic charges and thus can improve the specific capacitances' 10- 100 times of the electrode materials. Although a huge number of articles on supercapacitors are published and are increasing steadily each year, the electrochemical mechanisms of supercapacitors have not been fully understood. This fact resulted in difficulty in following the research directions on supercapacitors in general as well as the development of research on electrode materials for supercapacitors. Hence, the development of electrode materials for supercapacitors is extremely crucial [1].

Improvement in electrode materials of supercapacitors can be directly linked to the increased application of these devices. Batteries are the most common type of energy storage device and usually the most popular choice for technology-related applications. Batteries popularity is from the fact that they can provide a very high value of energy density and have the ability to store a large amount of energy in limited volume and weight. Supercapacitors bridge the gap between batteries (which are high energy density devices) and conventional capacitors (which are high power density devices). Supercapacitors have a much higher power density compared to batteries and have better shelf and cycle life but still there use is limited as they lack energy density. Research on testing new materials for electrodes serves the purpose of improving the energy density of supercapacitors, so the penetration of these devices in the energy storage market can be deepened [12]. However, supercapacitors are still finding several different commercial applications. Supercapacitors are used in different automobiles such as hybrid electric vehicles, electric vehicles, passenger trains and buses, etc use of supercapacitors in automobile sectors owes to the rapid charge and discharge. Electronic devices and energy harvesting systems like Wind turbines are some other application areas where supercapacitors are used [13].

Supercapacitors have several main advantages, such as having high power density because of their fast charge-discharge rates compared to traditional batteries. Also, supercapacitors' energy density is relatively higher than traditional dielectric capacitors because of their large surface area of porous electrode materials. Supercapacitors offer superior performance at a wide range of temperatures (between -40 °C to 70 °C) and provide high cyclability (> 100,000 cycles).

Supercapacitors can be divided into three main categories concerning their distinct charge storage mechanisms: i) Electrical Double-Layer Capacitors (EDLC) store charges directly at the electrode-electrolyte interface by the constitution of electrical double layers via a reversible non-Faradaic mechanism, ii) pseudocapacitors where the charge storage is made via reversible Faradaic charge transfer (i.e., surface redox reactions), and iii) hybrid supercapacitors in which charge storage mechanisms predicated on the augmentation of two previous mechanisms, i.e., Faradaic and non-Faradaic, into one device.

In EDLCs, the energy storage performance depends on the electrostatic interactions of charged particles at the electrode (mostly carbonaceous materials) and the electrolyte interface. The charge separation takes place in a Helmholtz double layer. On the other hand, in pseudocapacitors, the charge accumulation directly results from fast Faradaic redox reactions or intercalation processes occurring at (or near) the electrodes' surface, which are mostly made of state of art novel materials such as doped or undoped metal oxides, nanoceramics, and conducting polymers. One can combine these two principles into a single device by developing a hybrid (or so-called battery-type) supercapacitor, which exhibits both Faradaic and non-Faradaic properties with different configurations [1-3]. Hybrid supercapacitors can operate at high current densities, provide higher energy density at low current densities, and have a broader operating voltage range and higher upper limit voltage. For hybrid supercapacitors consisting of one capacitive and one faradaic (battery-like) electrodes, four-electrode configurations are possible: 1) battery-type positive and EDLC negative, 2) EDLC positive and battery-type negative, 3) pseudocapacitive positive and battery-type negative, and 4) battery-type positive and pseudocapacitive negative. All four different electrode configurations are also possible for asymmetric supercapacitors: 1) pseudocapacitive positive and EDLC negative, 2) EDLC positive and pseudocapacitive negative, 3) pseudocapacitive positive (X) and pseudocapacitive negative (Y), 4) EDLC positive (X) and EDLC negative (Y), where X and Y are different materials [4]. As a result, unlike battery research, it is not very common to name the electrode materials as anode or cathode in supercapacitors. The main reason for this is that the working mechanism of supercapacitors is not necessarily involved in ion insertion but rather absorption by Coulombic forces. Energy storage performances and capacitance of supercapacitors are affected by the Helmholtz layer thickness, presence of points defects in the electrode materials' microstructure, synthesis method, size, and morphology of the electrode material electrolyte type [5].

Most commercially available supercapacitors have a symmetrical configuration and consist of an anhydrous or organic electrolyte and generally use carbons with high surface area as active electrode material. The toxicity of these materials is one of the disadvantages of these existing supercapacitors [6]. Therefore, the constitution of environmentally friendly supercapacitors that have high power and energy capacity is of particular interest due to environmental pollution concerns.

In general, materials to be used as electrodes should have a significantly high surface area and sufficient pore distribution [7]. Thanks to developing technologies, especially nanotechnology, it has become easier to synthesize new alternative or novel materials having improved properties. Two main factors that need to be considered when developing environmentally friendly supercapacitors are following: obtaining electrodes in an environmentally friendly process and using water-based or non-toxic electrolytes.

A broad range of materials can be used as an electrode, such as metal oxides, carbons, conductive polymers, and composites [6]. Activated carbons (ACs) are amongst the highly used electrode materials to store electrical charges with an EDLC and a Faradaic mechanism. Bio-derived ACs can be synthesized from biomasses, which are agricultural wastes, thus producing electrode materials that are both low-cost and environmentally friendly [8]. Conductive polymers are one of the electrode candidates that can exhibit an environmental-friendly feature with the lack of heavy metals. Metal oxide-based pseudocapacitors have attracted significant attention due to their more stable electrochemical properties than conductive polymers. Also, they have comparatively higher capacitances and energy densities than carbon materials [6]. Noble metal oxides, such as $RuO_2$ and $IrO_2$, are some of the best transition metal oxides with excellent power densities and capacitive properties. However, unlike base metal oxides such as $MnO_2$, NiO, $Fe_3O_4$, etc., both $RuO_2$ and $IrO_2$ are not environmentally friendly [9, 10]. Manganese oxides and nickel oxides are highly promising due to their environmentally friendly behavior and high theoretical capacitances amongst base metal oxides [6, 11]. $Fe_2O_3$ and $Fe_3O_4$ can also be used as electrode materials for pseudocapacitors due to their minimal environmental damage. Since $Fe_2O_3$ has poor electronic conductivity, it is not suitable for energy storage devices that seek high power. In addition to this, $Fe_3O_4$ exhibits pseudocapacitance in alkali sulfide and sulfate aqueous medium [12]. Metal oxide nanoparticles such as CuZnCdO can also be used as non-hazardous and eco-friendly electrodes [13]. MXenes are the newest addition in electrode materials for supercapacitors application, which are gaining popularity in recent times. MXenes are 2D materials that are comprised of transition metal carbides, carbonitrides, and nitrides. MXenes are prepared by selective etching of element A from MAX phase precursors, where M represents an early transition metal, A is the III-A and IV-A element and X is carbon or nitrogen. MXene's increasing popularity as an electrode material is due to its excellent electrical and electrochemical properties such as hydrophilicity and surface area which eventually leads to the high-performance supercapacitor [2-4]. Polyoxometalates (POM) is a new class of electrode material that is now used in different energy storage devices. POMs can be used as building blocks for energy storage devices due to their nanometric oxide clusters with reversible redox activities [5]. POM stable molecular clusters give strong and reliable interaction with various electrode materials, so they are also tested in different supercapacitor applications [6,7]. Metal-organic frameworks (MOFs) are another comparatively new class of materials that are used in supercapacitors, their use in supercapacitors is due to their controlled pore size and their ability to adjust redox-metal centers. MOFs first used as electrode material was occurred in 2011 [8] and after that, there are continuous advances in testing this material in supercapacitors [9-11].

Selection of the proper electrolyte material is another significant step for supercapacitors. Electrolyte materials in all types of supercapacitors can be organic (e.g. acetonitrile, propylene carbonate), aqueous (e.g. KOH, NaOH, $H_2SO_4$), ionic liquid (IL) or solid. Organic electrolytes are relatively environmentally hazardous compared to aqueous electrolytes because they involve dissolving inorganic salts in organic solvents. Another electrolyte used in supercapacitors is solvent-free ionic liquids, which are eco-friendly because of their discardable vapor pressure. However, they are challenging to be considered commercial electrolyte candidates high-cost and non-biodegradability [14].

Moreover, for obtaining environmentally non-hazardous supercapacitors, biopolymers can be used as redox-active materials. Chitosan, chitin, natural polysaccharides, cellulose, and alginates are examples of biopolymers used to develop bioelectrodes. It needs to be clarified that these polymers are generally not used alone in supercapacitors but as composites with carbon-based materials and/or metal oxides [15]. Recently biocompatible electrode materials such as octacalcium phosphate have been used as an electrode to produce biosupercap and revealed high electrochemical performance [16]. However, the electrolyte type and nature must be selected by considering the nature of electrode materials, charge storage mechanism, and working voltage window of each supercapacitor.

In this review article, we have the ambition to discuss the supercapacitors and the logic behind their designs. We aimed to present a pathway for the reader to understand supercapacitors' design and the importance of the material selection. This paper also presents 10 supercapacitors with ultra-high specific and points out material selection significance.

## 2. 10-Best Ultra-High Specific Capacitance Supercapacitor Designs

In the study of Andikaey et al. [17] graphene nanocomposites that were coated with Nickel cobalt (NiCo) metal-organic framework was synthesized and utilized as electrode material for high-performance supercapacitor devices. The bimetallic organic framework NiCo-MOFs produced using the solvothermal method which provides improved electrical conductivity and better structural stability. Glucose-modified (GM) high-quality graphene (HQG) was manufactured here to be used as a platform for the growth and nucleation of bimetallic metal-organic framework nanosheets (NiCo/MOF). Also, to improve the electrochemical properties, MOFs are combined with liquid-phase exfoliated graphene (LEG). Via glucose modification, the processability of the LEG is increased and more active sites are provided for Ni and Co precursors to connect. The synthesized GM-LEG@NiCo-MOF nanocomposite exhibits an outstanding specific capacitance ($C_m$) value of about 4077 F/g at a current density (CD) of 2.5 A/g. The developed system provides a functional group that includes -COOH and -OH that offer active sites for the joining of Co and Ni precursors, this becomes the reason for very high capacitance value. The electrical conductivity and synergistic effect of GM-LEG@NiCo-MOF are better than the NiCo-MOF(pure) and LEG-based composite. Moreover, an asymmetric supercapacitor (ASC) is established where the developed nanocomposite and active carbon were used as electrodes and 3M KOH electrolyte is used in which a specific capacitance of 244 F/g and an energy density of 76.3 Wh/g is measured. Thanks to the unique speed ability, excellent cycle life, and excellent power and energy density as well as reversible capacitance, the GM-LEG@NiCo-MOF composites are suitable for future ASC applications.

Zardkhoshoui *et al.* [18] prepared nickel-cobalt oxides/hydroxides incorporated with reduced graphene oxide nickel foam (Ni,CoOH-rGO/NF) and used it as electrode material, the electrode was developed using a one-step simple and cheap electrodeposition method. The developed electrode showed ultrahigh specific capacitance of 4047 F/g at a current density of 3 A/g. The extremely high value of capacitance of Ni,CoOH-rGO/NF is due to the graphene nanosheets which are highly conductive and provide very fast electron transfer and secondly the size that is very small of Ni,CoOH on graphene sheets which provide material the chance to get completely used during the process. Moreover, using Ni,CoOH-rGO/NF electrode both symmetric supercapacitor (SSC) and an asymmetric supercapacitor (ASC) are assembled, the SSC was manufactured by the use of two identical sizes Ni,Co-OH/rGO/NF electrodes in PVA/KOH gel electrolyte. The ASC was assembled with Ni,Co-OH/rGO as the positive electrode, and FeS$_2$-rGO as the negative electrode. The energy density of 106.5 Wh/kg and 144 Wh/kg is measured from the SSC and ASC respectively. . According to the results, this paper proposes a way of constructing all-solid-state flexible energy storage devices to be used in the future. Tian *et al* [19] in this work used cobalt sulfide (CoS$_2$) as electrode material to be used in a high-performing supercapacitor. The electrode material used was prepared by two different synthesis techniques, which is the hydrothermal synthesis of cobalt sulfide on nickel foam (CoS$_2$@Ni) and surface generated flower-like nickel-iron layered double hydroxide (NiFe-LDH) nanospheres, are reported to design a binder-free electrode. The synthesized NiFe-LDH@CoS$_2$@Ni electrode shows a phenomenal specific capacitance of 3880 F/g at a current density of 1.17 A/g. The developed electrode was eventually used as in supercapacitor application, the assembled ASC was made by NiFe-LDH@CoS2@Ni hybrid used as a positive electrode and active carbon covered Ni foam utilized as the negative electrode and KOH/PVA hydrogel used as the electrolyte. Here, PVA acted both as a solid gel electrolyte and a separator simultaneously, this setup of KOH/PVA as electrolyte and separator provides remarkable safety and stability to the ASC. Moreover, an all-solid-state ASC, which exhibits an energy density of 15.84 Wh/kg which may be compared as the energy enough to light up a blue LED indicator, is assembled. Thanks to the flower-like ultra-thin nanosheet structure, which has a large surface area and a large number of active sites, the diffusion distance between electrolyte ions and the substrate is decreased which will result in fast electron transport and proved the large potential of NiFe-LDH@$CoS_2$@Ni in high-performance supercapacitor applications.

In the work of Kumbhar *et al* [20], a novel nickel cobaltite nickel molybdate ($NiCo_2O_4$/$NiMoO_4$) nanoflake/nanoparticle core/shell electrodes on nickel foam (NF) are fabricated and used in high energy supercapacitor application. The electrode materials were prepared using a binder-free, cheap, and simple electrodeposition technique. The developed electrode material showed an ultra-high specific capacitance of 3705 F/g at 1.5 A/g and has a %94.6 cycle stability after 5000 charges/discharge cycles. The nanoporous heterostructure of $NiCoO_2$ nanoflakes and $NiMoO_4$ nanoparticles leads to a fast ion transfer, increase in electrical conductivity, a decrease in diffusion length for ions, and the synergy between the nanoflakes and nanoparticles offers a large active surface area important for electrochemical energy storage and oxygen evolution electrocatalysis which results in such high capacitance value. Finally, an SSC was assembled by $NiCo_2O_4$/$NiMoO_4$ electrodes, a copper tape was applied for electrical contacts, a piece of cellulose filter was utilized as a separator and 3M KOH electrolyte was used. The assembled SSC exhibits an energy density of 76.45 Wh/kg and has %88 cycle stability after 5000 cycles. The remarkable performance of the device suggests that the $NiCoO_2$/ $NiMoO_4$ core/shell heterostructure can be synthesized using the strategy in this paper as an electrode material and can be used for energy storage devices in future applications.

In the study of Xu *et al.* [21], the fabrication of manganese molybdate hydrates manganese oxide ($MnMoO_4H_2O$@$MnO_2$) core-shell nanosheet arrays on nickel foam is reported which was then used as electrode material in supercapacitor. $MnMoO_4H_2O@MnO_2$ is synthesized using a simple two-step hydrothermal process in absence of a calcined process taking place at high temperatures. Three $MnMoO_4H_2O@MnO_2$ are synthesized in reaction times 1h, 3h, and 6h and called MMM1, MMM3, MMM6 respectively. The MMM3 exhibits an ultra-high specific capacitance of 3560 F/g at 1 A/g compared to MMM1 and MMM6 and offers %841 cycle stability after 10,000 cycles. It is seen that the size of the $MnO_2$ shell electrode is very important to have a good specific capacitance because the MMM1 was too sparse and MMM6 was too thick compared to MMM3. $MnMoO_4H_2O@MnO_2$ core-shell nanosheet arrays are far more effective in providing electrochemical activity compared to a single $MnMoO_4H_2O$ electrode. The developed electrode offers a large surface area and more number of electrochemically active sites which gives cycling stability and increases capacitance, it provides rapid ion diffusion and electron transport at electrode/electrolyte interface and synergistic effect of $MnMoO_4H2O$ and $MnO_2$ which improves the performance of the supercapacitor. The ASC was manufactured by utilizing activated carbon (AC) on the platform of nickel foam as the negative terminal of the electrode and $MnMoO_4H2O@MnO_2$ nanosheet arrays on Ni foam as the positive electrode while PVA/KOH used as electrolyte with cellulose paper as the separator. The supercapacitor developed exhibits 45.6 Wh/kg energy density at a power density of 507.3 Wh/kg. The electrode system assembled proves that the $MnMoO_4H_2O@MnO_2$ is suitable to use in future energy storage applications.

Xing *et al.* [22] reported a smart strategy to combine nickel phosphide (Ni-P) with nickel-cobalt layered double hydroxide (NiCo- LDH) to fabricate core-shell nanorod arrays (NRAs) on NF which was used as an electrode for application in supercapacitor devices. NiCo-LDH nanosheets were developed on the surface of Ni-P nanorods using an easy electrodeposition method. The fabricated Ni-P@NiCo LDH electrode gives an ultra-high specific capacitance of 3470 F/g at 1.3 A/g owing mostly to the better stability and electrochemical activity provided by the NiCo-LDH, abundancy of interface which the core-shell nanorods offer, and the hierarchical structure of 1D nanorods, 2D nanosheets, and the 3D Ni foam. The asymmetric supercapacitors (ASCs) was manufactured using Ni-P@NiCo LDH as the positive side of the electrode and activated carbon (AC) as the negative terminal of the electrode, one piece of polypropylene fiber was used in between the electrode which serves as the separator, and 6M KOH was electrolyte in the supercapacitor design. The assembled ASC showed 35 Wh/kg energy density at a power density of 770.8 Wh/kg, the attained energy density can be used to light an LED for around 40 mins. The obtained result is very promising to use this electrode system for future supercapacitor applications.

In the study of Cheng *et al* [23], assembly of the cobalt oxide ($CoO_x$) mesoporous microspheres with $Co_3O_4$ as core@shell model ($CoO_x@Co_3O_4$) using the solvothermal route and post-annealing treatment with $NaH_2PO_4 \cdot 2H_2O$, is aimed. The synergy between $CoO_x$ and $Co_3O_4$ results in $CoO_x$ mesoporous microspheres present an ultra-high specific capacitance of 3377 F/g at a current density of 2 A/g, which is relatively higher than the individual specific capacitance of CoO and $Co_3O_4$ obtained from 3-point techniques. The phenomenal performance of this system is credited to the fact that both $Co_3O_4$ and CoO have very high theoretical capacitance, the three-dimensional structure, and high number mesopores are assisted for high surface accession of $OH^-$ ions, the synergistic effect of CoO and $Co_3O_4$ offered by the core-shell structure due to common redox reaction and the scobinate surface which is the small nanoparticles contained in the microspheres. Furthermore, an ASC compose of $CoO_x$ and graphene as its electrode and 6M KOH is used as the electrolyte, which exhibits a specific capacitance of 123 F/g at a current density of 1 A/g and energy density of 44.06 Wh/kg, is assembled. Due to the high specific capacitance, decent cycle stability, and being reversible of the faradaic reaction of the assembled ASC, $CoO_x$ mesoporous microspheres are seen as a promising material.

Nickel hydroxide/Carbon nanotube ($Ni(OH)_2$/CNT) electrodes were synthesized and used in a supercapacitor device by Tang *et al* [24]. Carbon nanotube (CNT) bundles are grown on the NFs by a chemical vapor deposition (CVD) technique and subsequently nickel hydroxide $Ni(OH)_2$ deposited onto the CNT layers which results in the fabrication of $Ni(OH)_2$/CNT/NF electrode that exhibits a specific capacitance of up to 3300 F/g. This very high value of capacitance of the developed electrode can be considered due to the thorough consumption of the well-dispersed $Ni(OH)_2$ nano-flakes on the CNT/NF current collector which offers a high surface area. Additionally, the conductive nature of CNT and the disorder characteristic of $Ni(OH)_2$ structure which occurred from intercalation of anions and water molecules increased the active surface area further, with the increase in carrier transport as well. Furthermore, an ASC is assembled in which the Ni $(OH)_2$/CNT/NF electrode is used as anode and Active Carbon (AC) is used as cathode and delivers a 50.6 Wh/kg energy density using KOH as an electrolyte. Thanks to the high specific capacitance value combined with cycling stability, assembled ASC is seen as a potential candidate for future energy storage devices.

Carbon/Cobalt oxide(CoO) nanostructured arrays were synthesized for enhancing asymmetric supercapacitor device performance by Wang *et al.* [25]. The hybrid nanowire arrays (NWAs) were synthesized by the growth of high electrically conductive carbon onto the rough CoO NW arrays on 3D NF. The CoO@C nanostructure arrays (CCNAs) were fabricated by hydrothermal routes after which controlled annealing and carbon deposition process at low temperature in CVD was performed, and an extremely high specific capacitance value of 3282 F/g is obtained. As both the component in the electrode system are electrically active so they enhance the electrical conductivity, the electrochemical ions were then more convenient to absorb on the surface of active material which results in improved redox reaction which increases the capacitance more. Moreover, an ASC, in which CCNAs are used as the positive terminal of electrode and AC as the negative terminal of the electrode with the electrolyte solution of 6M KOH and one piece of cellulose paper as the separator is assembled and exhibits the highest energy density (as 58.9 Wh/kg at a power density of 198.7 W/kg) up to now amongst the Co-based supercapacitors. Moreover, this study proposes both the fabrication of graphite carbon/ cobalt monoxide nanostructure arrays and the possibility of the fabrication of 1D graphene arrays.

In the study of Zhong *et al* [26], nickel cobalt manganese ternary carbonate hydroxide is used as a novel electrode material for improving supercapacitor performance. (NiCoMn-CH) ultra-thin nanoflakes are coated on cobalt carbonate hydroxides (Co-CH) NWAs on top of an NF using a two-step scalable solvothermal method in the absence of any binder and surfactant. The fabricated Co-CH@NiCoMn-CH electrode is a novel electrode due to its high rate and long cycle life and presents a high capacitance of 3224 F/g at a current density of 1 A/g using a 3-electrode system. The electrode also shows an outstanding cycling performance and reaches 92.4% of the initial specific capacitance after the 6000 cycles. The methodology used to develop this electrode system not only prevent the typical aggregation and give sufficient ion transfer but also reduce the charge transfer of developed Co-CH subsequently leading to enhanced electrical conductivity for electron transfer, which is the reason for such a high value of capacitance. Furthermore, an ASC, in which Co-CH@NiCoMn-CH is used as the positive electrode and activated carbon is used as the negative electrode, is assembled and that exhibits a 20.31 Wh/kg energy density at a power density of 748.46 Wh/kg at a 0-1.5 V voltage window. It should be noted that using the assembled ASC, a blue LED is lightened brightly for more than 90 seconds and this accomplishment proves the viability of this study and the importance of the selected strategy to fabricate this material at low cost and temperature as well as its strategic design and will lead this material to be a promising candidate in the future.

In summary, all electrochemical performance parameters for the supercapacitor designs given above have been listed in Table 1.

Table 1: Selected Supercapacitors design and their respective properties

|    | Electrode Materials | Electrolyte | Specific Capacitance ($C_m$) (F/g) | Current Density (j) | Energy Density (Wh/g) | Ref. |
|---|---|---|---|---|---|---|
| 1. | GM-LEG@NiCo-MOF/Active Carbon | 3M KOH solution | 4077 | 2.5 A/g | 76.3 | [17] |
| 2. | Ni,Co-OH/rGO /$FeS_2$-rGO | PVA/KOH gel | 4047 | 3 A/g | 144 | [18] |
| 3. | NiFe-LDH@$CoS_2$@Ni/Active Carbon | KOH/PVA hydrogel | 3880 | 1.17 A/g | 15.84 | [19] |
| 4. | $NiCo_2O_4$/$NiMoO_4$ | 3M KOH solution | 3705 | 1.5 A/g | 76.45 | [20] |
| 5. | $MnMoO_4H_2O$@$MnO_2$/Active Carbon | PVA/KOH gel | 3560 | 1 A/g | 45.6 | [21] |
| 6. | Ni-P@NiCo LDH/ Active Carbon | 6M KOH solution | 3470 | 1.3 A/g | 35 | [22] |
| 7. | CoOx@$Co_3O_4$/Graphene | 6M KOH | 3377 | 2 A/g | 44.06 | [23] |
| 8. | Ni$(OH)_2$-CNT-NF/Active Carbon | KOH solution | 3300 | NA | 50.6 | [24] |
| 9. | CCNAs/Active Carbon | 6M KOH solution | 3282 | NA | 58.9 | [25] |
| 10. | Co-CH@NiCoMn-CH/Active Carbon | NA | 3224 | 1 A/g | 20.31 | [26] |

## 3. Conclusions

Here, we basically discuss supercapacitors, their types, design parameters, and the importance of material selection and gave 10 different ultra-high specific capacitance supercapacitors pointing out their particular high capacitance in terms of materials science. 10 different innovative approaches and significant outcomes presented in this review make the combination of components with supercapacitors an interesting direction for developing novel, high-performance, and practically useable electrochemical energy storage technologies. It should be noted that special care must be given to provide a synergy of the four main parts of supercapacitors: positive and negative electrodes, electrolyte, and separator to attain very high capacitance values. Optimizing the performance of the material is vital for elevating the performance, especially capacitance and cyclability of supercapacitors. Instead, performance evaluation and comparison between supercapacitors should be based on power capability and, energy capacity and should also take into account the mass of all components. However, in most of the supercapacitor research priority has been given to the specific capacitance values compared to energy density. From these 10 ultra-high capacitance supercapacitor designs, one may conclude that scientists try to achieve high performance and give high priority to clean, environmentally friendly, and sustainable materials choices. To develop new materials with optimal performance, three important research directions in supercapacitors are trending: electrodes from composites and nanomaterials, polymeric electrolytes. Overall, the immediate goal should be to research more and more on the correlation between the properties of the electrolytes and the electrodes so that a better understanding of the working principles and designs can be achieved and used to guide scientists for further improvement and engineers for better production.

**Acknowledgments**

This study is supported by a research grant from the Scientific and Technological Research Council of Turkey (TUBITAK, Grant No: 118C243) in the frame of 2232-International Fellowship for Outstanding Researchers. Partially supported by the Starting Grant of Sabanci University with the grant number: B.A.CF-19-01962.

21